\documentclass[twocolumn,preprintnumbers,amsmath,amssymb]{revtex4}

\usepackage{amsmath}
\usepackage{graphicx}
\usepackage{dcolumn}
\usepackage{bm}
\begin{document}

\title{The effect of self-affine fractal roughness of wires on atom chips }

\author{Z. Moktadir}
 \altaffiliation[]{School of Electronics and Computer Science, Southampton
University, UK.}
\author{B. Darqui\'{e}}
 \altaffiliation[]{Centre for Cold Matter, Blackettt Laboratory, Imperial College London,
UK}
\author{M. Kraft}
 \altaffiliation[]{School of Electronics and Computer Science, Southampton
University, UK.}
\author{E.A. Hinds}
 \altaffiliation[]{Centre for Cold Matter, Blackettt Laboratory, Imperial College London,
UK}

\begin{abstract}
Atom chips use current flowing in lithographically patterned wires
to produce microscopic magnetic traps for atoms.  The density
distribution of a trapped cold atom cloud reveals disorder in the
trapping potential, which results from meandering current flow in
the wire. Roughness in the edges of the wire is usually the main
cause of this behaviour. Here, we point out that the edges of
microfabricated wires normally exhibit self-affine roughness. We
investigate the consequences of this for disorder in atom traps. In
particular, we consider how closely the trap can approach the wire
when there is a maximum allowable strength of the disorder. We
comment on the role of roughness in future atom--surface interaction
experiments.
\end{abstract}

\maketitle

\section{Introduction}
Atom chips are microfabricated structures that allow the preparation
and manipulation of cold atom clouds or Bose--Einstein condensates
(BEC) above a substrate surface. Often, these structures use
current-carrying wires to produce tightly confining magnetic
microtraps close to the substrate surface, where atom clouds can be
held still, transported, or
split~\cite{Hinds2001,Fortagh98,Denschlag99,Key2000,Fortagh07}. With
the integration of optical components~\cite{Eriksson05} and movable
structures~\cite{Gollasch05} into atom chips, new possibilities are
now opening for neutral atoms  on a chip to form quantum sensors,
clocks and information processors~\cite{Calarco04}. However, the
homogeneity and stability of atom clouds can be compromised close to
a metallic surface by physical factors that cause fragmentation
and/or the loss of atoms. Two main phenomena have been identified:
(i) spatial imperfections of the wire, which cause the current to
flow non-uniformly and make the atom trap rough, and (ii) thermal
fluctuations of the magnetic field near the surface, which drive
spin flips of the atoms and cause loss~\cite{Jones03}. The first of
these is the subject of our paper.

Recently, the corrugation of magnetic fields close to a wire has
been studied extensively. Initial experiments showed that atom
clouds break up into fragments as they approach the
surface~\cite{Zimmermann1,Jones04}, then it was  demonstrated that
this is due to a magnetic field component parallel to the
wire~\cite{Zimmermann2}, caused by transverse components of the
current density. Some theoretical efforts were made to relate this
to the details of the current flow~\cite{Jones04} and to roughness
of the surface and irregularity in the edges of the
wire~\cite{Wang04,Schumm05}, which cause the current to meander. If
the meander has a single spatial Fourier component of wavevector
$q_0$, the decay of this anomalous field decreases with distance $d$
above the wire according to the Bessel function $K_1(d q_0)$ to a
good approximation~\cite{Jones04}. When the transverse current has a
broad noise spectrum, this decay can sometimes be described by a
power law~\cite{Zimmermann2}.

Conductors lithographically patterned on an atom chip are usually
fabricated with good bulk homogeneity in order to minimise this
potential problem of magnetic roughness. The width of the wire is
typically comparable to the distance between the magnetic trap and
the surface. In these typical cases, the meandering of the current
is driven mainly by the roughness of the edges, as determined by the
fabrication process. Three processes are available, namely, (i)
electrodeposition of the metal into a mould formed by a thick
photoresist~\cite{Kraft04,Schumm05}, (ii) etching of a complete
metallic film to create the space between wires~\cite{Kraft04} using
wet chemicals or ion beam milling~\cite{Lewis06}, and (iii)
evaporation of the metal onto the substrate through a mask formed by
a patterned resist (the method known as lift-off).

With all these methods of microfabrication, the edges of the wires
exhibit self-affine fractal roughness~\cite{Meakin,Barabasi}, a type
of roughness that we now discuss. Consider an edge along the $z$
direction with roughness fluctuations $f(z)$. By definition the
height-height correlation function is given by
$G(r)^2=\langle[f(z)-f(z+r)]^2\rangle$, the autocorrelation function
is $C(r)=\langle f(z)f(z+r)\rangle$, and the mean square roughness
is $\sigma^2=\langle f(z)^2\rangle$. The angle brackets denote
averaging over the (large) length of the wire. These quantities are
connected by the relation $G(r)^2=2\sigma^2-2C(r)$. A self-affine
fractal edge is one that satisfies the scaling law $G(r)\propto
r^\alpha$, where $\alpha$ is known as the roughness exponent or
Hurst exponent. The statistical properties of such an edge are
invariant when the length is scaled by a factor $\lambda$, provided
there is an accompanying scaling of the transverse dimension by
$\lambda^{\alpha}$. Microfabricated edges exhibit precisely this
type of behaviour on small length scales~\cite{Constantoudis03} up
to a characteristic length $\xi$, known as the correlation length.
For $r>\xi$, $G(r)$ tends to the constant value $\sqrt{2}\,\sigma$
and the autocorrelation function $C(r)$ tends to zero. This
behaviour is captured by the empirical autocorrelation function
\begin{equation}\label{correlation}
   C(r)=\sigma^2 \exp[-(r/\xi)^{2\alpha}],
\end{equation}
which has the required assymptotic behaviour at large and small $r$
and fits experimental data well~\cite{Palasantzas93}. The Hurst
exponent $\alpha$ is normally between 0 and 1~\cite{note1}, while
the correlation length and rms roughness are both typically in the
range 1--100~nm~\cite{Schumm05,Constantoudis03,Palasantzas93}.

Considerable progress has been made in understanding how roughness
in the edges of a wire can generate roughness in the magnetic traps
produced by the wire~\cite{Jones04,Wang04,Schumm05}. However, the
analyses to date have considered edges with a white noise spectrum
or other rather specific model spectra. Here we reconsider the
magnetic noise of atom traps, taking into account this realistic and
more generally applicable model of the self-affine fractal edge
roughness.

\section{Description of self-affine roughness along an edge\label{description}}
\begin{figure}
  \centering
  \includegraphics[width=5cm]{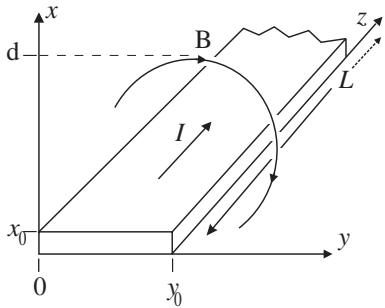}\\
  \caption{Sketch of the ideal wire geometry under consideration. Current $I$
    flows uniformly along $z$. At height $d$ above the wire, this makes a
    field along $y$. Roughness in the edges of the wire cause the current to
    deviate from side to side, producing a noise field $\delta B_z$.}\label{wire}
\end{figure}

\begin{figure}
  \centering
  \includegraphics[width=8cm]{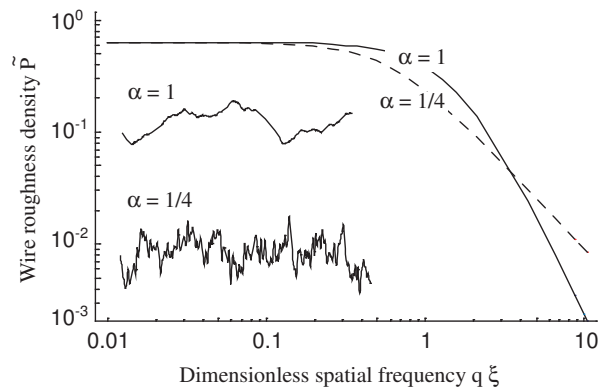}\\
  \caption{Spectrum of wire roughness $\tilde{P}$ defined
  by equation (\ref{powerspeca}) versus $q \xi$,
  where $\xi$ is the correlation length. Spectra are plotted for two Hurst exponents,
  $\alpha = 1/4$ and $\alpha = 1$. Inset are two representative
  plots showing the roughness over a length $2\xi$ along $z$
  with these two values of $\alpha$.}
  \label{Pspectrum}
\end{figure}

Figure \ref{wire} defines a coordinate system and shows the wire
that we are considering. The left and right edges of the wire lie on
$y=0$ and $y=y_0$ with fluctuations $f_L(z)$ and $f_R(z)$
respectively. Hence the centre of the wire lies on $y_0/2+\delta
y(z)$, where $\delta y(z)=\tfrac{1}{2}\left[f_L(z)+f_R(z)\right]$,
with correlation function $C(r)=\langle\delta y(z+r) \delta
y(z)\rangle$. We take as our starting point the empirical
correlation function $C(r)$ given in equation (\ref{correlation}).
In the particular case when $\alpha=1/2$, the corresponding power
spectrum has the Lorentzian form
\begin{equation}\label{powerspec}
    P(\textstyle\frac{1}{2}\displaystyle,q)=\textstyle\frac{2}{\pi}\displaystyle\Re e\int _0 ^{\infty} C(r) e^{-iqr} dr=\sigma^2 \xi\frac{2/
    \pi}{1+q^2\xi^2},
\end{equation}
where $q$ spans the range $0$ to $\infty$. In order to have
analytical results for a more general range of possibilities, we
extend the power spectrum of equation (\ref{powerspec}) to the form
\begin{equation}\label{powerspeca}
    P(\alpha,q)=\sigma^2\xi\frac{2/\pi}{(1+a
    q^2\xi^2)^{\frac{1}{2}+\alpha}}\equiv\sigma^2\xi \tilde{P}.
\end{equation}
This is a one-dimensional version of the approximation introduced by
Palasantzas (section IV of \cite{Palasantzas93}) to describe surface
noise. The parameter $a$ in the denominator of this spectrum is
needed to ensure that the integral of equation (\ref{powerspeca})
over all $q$ yields the mean square roughness $\sigma^2$. This
normalisation condition requires
\begin{equation}\label{a}
    a=\frac{\Gamma^2(\alpha)}{\pi \Gamma^2(\frac{1}{2}+\alpha)},
\end{equation}
where $\Gamma$ is the Euler function. We find by direct numerical
integration that the power spectrum corresponding to equation
(\ref{correlation}) is reasonably well reproduced by equation
(\ref{powerspeca}), but only over the range $1/4<\alpha<1$. Figure
\ref{Pspectrum} shows the dimensionless spectrum $\tilde{P}$,
defined in equation (\ref{powerspeca}), for these two extremes of
the Hurst exponent $\alpha$. At low frequency it has the value
$(2/\pi)(\sigma^2 \xi)$ regardless of $\alpha$, but as the frequency
increases, the spectrum with lower $\alpha$ also has lower noise. At
higher frequency still, this necessarily reverses because these
spectra are normalised. Typical profiles of the centre position
$y_0/2+\delta y(z)$, plotted over a length $2\xi$, are inset into
the figure to illustrate this.  The case of $\alpha=1/4$ exhibits
more high-frequency noise but less long-wavelength noise than that
of $\alpha=1$.

This model is expected to describe fluctuations in the centre of a
wire fabricated on an atom chip. In particular, it describes the
atom chip wires currently being used at Imperial College London,
which are made by ion beam milling a gold film. These typically have
roughness $\sigma\simeq 3$~nm, correlation length $\xi\simeq20$~nm
and Hurst exponent $\alpha \simeq 0.5$. The noise reported in figure
8 of Schumm et al. \cite{Schumm05} is also consistent with this
model, giving for their evaporated wire the values $\sigma\simeq
1.4$~nm, $\alpha \simeq 0.8$ and $\xi\simeq 50$~nm. (Note, however
that such analysis of the power spectrum is not a very reliable way
to measure $\xi$ or $\alpha$~\cite{Constantoudis04}). In the
following, we discuss the roughness of the magnetic atom traps
produced when current flows through such a wire and we investigate
how the field fluctuations vary with the Hurst exponent $\alpha$ and
correlation length $\xi$. Wires made by electrodeposition into a
thick photoresist mould have so far been much rougher. For example,
the electroplated wire of \cite{Schumm05} had $\sigma\simeq 70$~nm,
$\alpha \simeq 0.5$ and $\xi\simeq 200$~nm. Moreover, the spectrum
of that wire exhibited a second power-law region with exponent
$-2.2$ at wavelengths longer than $20~\mu$m, indicating a second
regime of correlated roughness.

\section{Roughness of the atom trap formed by a wire}

Consider the wire in figure \ref{wire}. If the current $I$ flows
uniformly along the $z$-direction, the magnetic field lines lie in
the $xy$ plane. In particular, the field points in the $y$-direction
above the centre of the wire. This field is cancelled at a height
$d$ by applying an opposite uniform bias field and the resulting
line of zero magnetic field is surrounded by a transverse quadrupole
field. Magnetic atoms can then be trapped at height $d$ above the
centre of the wire by the magnetic dipole interaction $ - \vec{\bm
\mu} \cdot \vec{{\bf B}}$. A small uniform bias field $B_z$ is often
applied along the $z$-direction as well, so that the magnetic field
minimum goes to $B_z$ rather than to zero. This suppresses the loss
of atoms through non-adiabatic spin flips.

\begin{figure}
  \centering
  \includegraphics[width=8cm]{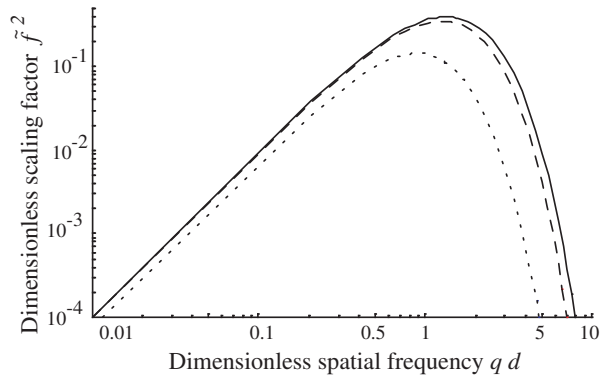}\\
  \caption{Scaling factor $\tilde{f}^2$ as a function of
  the dimensionless frequency $q d$. This quantity links the noise in
  the edges of the wire to the noise in the magnetic atom-trapping potential.
  Curves are shown for three ratios of height to width: $d/y_0=10$ (solid curve),
  2 (dashed) and 0.6 (dotted).}\label{fspectrum}
\end{figure}

In reality, the noise in the edges of the wire causes the current to
deviate from side to side, generating a noise field component
$\delta B_z$ along the $z$ direction. Consequently, the potential
energy along the centreline of the trap is no longer the constant $
- \mu_z B_z$ but is modulated by noise $ - \mu_z \delta B_z$.
Assuming that the wire is thin (along x) compared with the height
$d$, the power spectrum of this noise can be written as
\begin{equation}\label{Sspectrum_eq}
    S(q)=\mu_z^2 B_0^2 \left(\frac{\sigma^2\xi}{d^2}\right)\tilde{P}\tilde{f}^2
    =\mu_z^2 B_0^2 \left(\frac{\sigma^2\xi}{d^2}\right)\tilde{S},
\end{equation}
where $B_0=\mu_0 I/2 \pi d$ is characteristic of the ideal field
produced by the wire and $\sigma^2 \xi \tilde{P}$ is the power
spectrum describing fluctuations in the centre of the wire, which we
take here to be given by equation (\ref{powerspeca}). The
dimensionless scaling factor $\tilde{f}^2$ translates the noise in
the centre of the wire to the noise in the field. It is given by
\cite{Wang04}
\begin{eqnarray}
  \tilde{f} &=& (q d)^2\,\frac {2\sinh(\textstyle\frac{1}{2}q y_0)}{q y_0\sinh(q y_0)}
  \sum_{n=0}^{\infty}\frac{(-1)^n K_{n+1}(q d)}{n!(2 q d)^n}\nonumber \\
  &  &\times [\gamma_{2n+1}(\textstyle\frac{1}{2}q y_0)-\gamma_{2n+1}(-\textstyle\frac{1}{2}q y_0)],
\end{eqnarray}
where $K_n(x)$ is the modified Bessel function of the second kind
and $\gamma_n(x)$ is the incomplete Gamma function. This expansion
is useful in the range $d>y_0/2$, where a small number of terms is
sufficient to achieve convergence: 50 terms at $d=0.6 y_0$ and fewer
terms at larger distance. When $d<y_0/2$, the individual terms
become excessively large and the series appears not to converge. In
summary, the spectrum $S(q)$ of the noise in the magnetic atom trap
depends on the Hurst exponent $\alpha$ and four length scales:
$\sigma$ and $\xi$ in $P(\alpha,q)$, which characterise the
roughness of the edges and $y_0$ and $d$ in $\tilde{f}$, which
define the geometry of the trap. The over-all energy scale is given
by $\mu_z B_0$.

The frequency dependence of $\tilde{f}^2$ is illustrated in figure
\ref{fspectrum}. The three curves correspond to wire widths of
$d/10$ (solid line), $d/2$ (dashed) and $d/0.6$ (dotted). When $y_0$
is small compared with $d$, $\tilde{f}$ is quite insensitive to its
value, but $\tilde{f}$ becomes small for a wider wire as the edges
move further away compared with $d$. At low frequencies, i.e. when
$q d <<1$, the function \mbox{$\tilde{f}^2\backsimeq[q
d\,(\frac{2d}{y_0})\arctan({\frac{y_0}{2d}})]^2$} increases in
proportion to $q^2$, reaching a maximum in the vicinity of $q d=1$.
For large $q d$, the function decays as
\mbox{$\tfrac{\pi}{2}\,qd\,(\tfrac{2d}{y_0})^2\exp\{-qd\,[2+(\tfrac{y_0}{2d})^2]\}$}.
If the current wanders periodically from side to side with a given
amplitude, the angular variation of the current density $\mathbf{j}$
is inversely proportional to the wavelength. Consequently, the
transverse component $j_y$ is proportional to the frequency $q$.
This is the physical cause of the linear cutoff in $\tilde{f}$ at
low-frequency. The exponential (Bessel) cutoff at high frequency is
due to Laplace's equation for the magnetostatic potential, which
naturally smoothes high frequency ripples as one goes far away from
the wire.

\begin{figure}
  \centering
  \includegraphics[width=8cm]{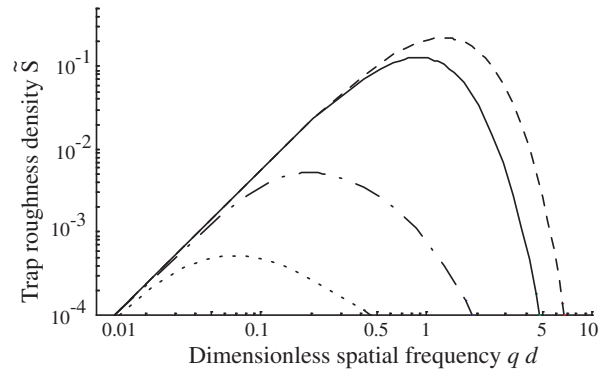}\\
  \caption{Plot of the product $\tilde{P}\tilde{f}^2$, which determines
   the roughness spectrum $\tilde{S}$ of the magnetic atom trap. The Hurst
   exponent is $\alpha=1$ and $d=2 y_0$. Curves are
   shown for four ratios of correlation length $\xi$ to distance $d$:
   $\xi /d=33$ (dotted line), 10 (dash-dotted line), 1 (solid line),
   0.01 (dashed line).} \label{Sspectrum_fig}
\end{figure}

The dimensionless spectrum $\tilde{S}$ of the noise in the trapping
potential (see equation (\ref{Sspectrum_eq})) is the product of the
two spectra $\tilde{P}$ and $\tilde{f}^2$, shown in figures
\ref{Pspectrum} and \ref{fspectrum}. Whereas $\tilde{P}$ depends on
the frequency through $q \xi$, $\tilde{f}$ is a function of $q d$,
therefore the shape of the spectrum $\tilde{S}$ depends on the ratio
$d/\xi$ as illustrated in figure \ref{Sspectrum_fig}.  The dashed
curve in figure \ref{Sspectrum_fig} represents the case of short
correlation length, $\xi=d/100$, for which $q\xi<<1$ over the whole
range of the graph, making $\tilde{P}$ constant at $=2/\pi$. In this
limit, equation (\ref{Sspectrum_eq}) gives $S(q)=\mu_z^2 B_0^2
\left(\frac{2\sigma^2\xi}{\pi d^2}\right) \tilde{f}^2$, a spectrum
that is independent of the wire roughness except for the $\sigma^2
\xi$ in the over-all scale factor.  This general behaviour of a
spectrum proportional to $\tilde{f}^2$ persists throughout the range
$\xi\lesssim d$, as also indicated by the solid line representing
$\xi=d$. At the other extreme, the dotted curve in figure
\ref{Sspectrum_fig} represents the case of long correlation length,
$\xi=33d$, for which $q d \ll 1$ over the whole range of interest,
giving $\tilde{f}^2\simeq(q d)^2$ and therefore $S(q)\simeq\mu_z^2
B_0^2 \,\sigma^2\xi \,q^2 \tilde{P}$. This general behaviour of
$S(q) \propto q^2 \tilde{P}$ is characteristic of the whole range
$\xi>d$, as also illustrated by the dash-dotted line in figure
\ref{Sspectrum_fig}.

\begin{figure}
  \centering
  \includegraphics[width=8cm]{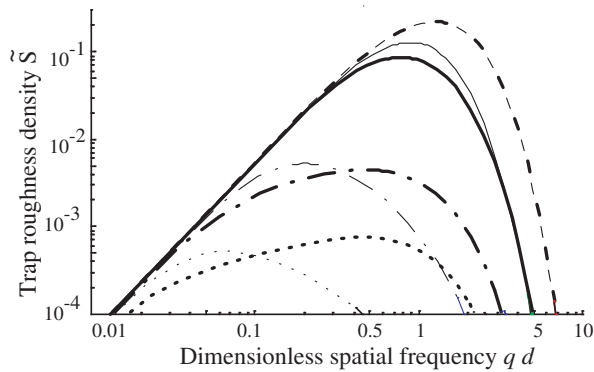}
  \caption{Roughness spectra $\tilde{S}$ of the magnetic trap for two
  values of the Hurst exponent $\alpha$. Light curves: $\alpha=1$.
  Heavy curves: $\alpha=1/4$. As in figure \ref{Sspectrum_fig},
  spectra are given for $\xi /d=33$ (dotted), 10 (dash-dotted), 1 (solid),
  and 0.01 (dashed).} \label{alphaSspectrum}
\end{figure}

Figure \ref{alphaSspectrum} shows the effect of changing the Hurst
exponent in the roughness spectrum of the wire from $\alpha=1$, as
in figure \ref{Sspectrum_fig} (light curves), to $\alpha=1/4$ (heavy
curves). When $\xi\ll d$ (dashed curves), the change of Hurst
exponent makes no difference because the spectrum is essentially
independent of $\tilde{P}$. By contrast, the dotted curves
representing $\xi=33d$ exhibit a strong dependence on the Hurst
exponent. Reducing $\alpha$ from 1 to 1/4 suppresses the low
frequency noise and increases the power at higher frequencies, as
already noted in the context of figure \ref{Pspectrum}. This has the
effect of moving the peak of the noise spectrum to higher
frequencies. The same effect is seen in the dash-dotted curves of
figure \ref{alphaSspectrum} representing $\xi=10d$. When $\xi$ is
equal to $d$ (solid curves), the change to $\alpha=1/4$ suppresses
the low frequency part of the spectrum, but the corresponding
increase at higher frequency is not evident because the spectrum is
cut off at higher frequencies by the exponential roll-off of the
function $\tilde{f}^2$.

\begin{figure}
\centering
\includegraphics[width=8cm]{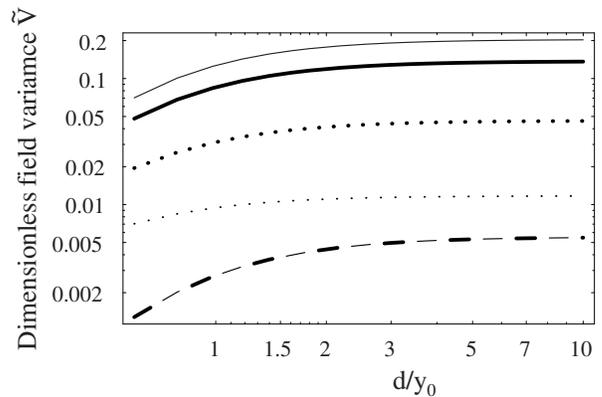}
\caption{Dimensionless magnetic field variance $\tilde{V}$ versus
distance $d$ of the trap from the wire, normalised to the width
$y_0$ of the wire. Curves are given for $\xi/d=1$ (solid), 20
(dotted) and 0.01 (dashed). Light curves: $\alpha=1$. Heavy curves:
$\alpha=0.25$.} \label{Vtilde_fig}
\end{figure}
\begin{figure}[b]
\centering
\includegraphics[width=8cm]{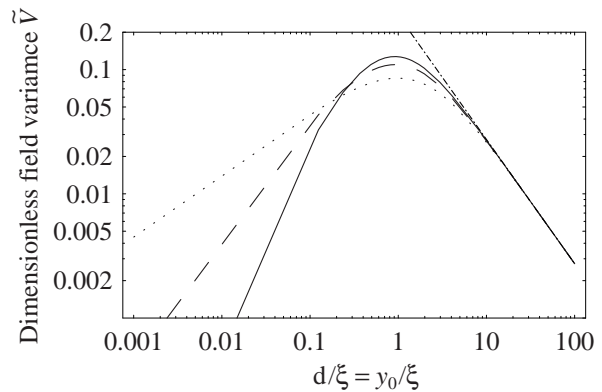}
\caption{Dimensionless magnetic field variance $\tilde{V}$ versus
the ratio $d/\xi$ for the particular case $d=y_0$. Solid line:
$\alpha=1$, dashed line: $\alpha=1/2$, dotted line: $\alpha=1/4$.
Dashed-dotted line: $\xi/d$ law given by equation
(\ref{Vtildelimit}) and valid when $d\gg\xi$.} \label{Vtildexi}
\end{figure}
The mean square roughness of $B_z$, let us call it $V$ for variance,
is obtained by integrating equation (\ref{Sspectrum_eq}) over
frequency:
\begin{equation}\label{V_eq}
    V(\alpha,\xi,d,y_0) \equiv \langle B_z^2 \rangle =
    \frac{1}{\mu_z^2}\int_0^{\infty}S(q)dq\,.
\end{equation}
Once again, it is useful to adopt a dimensionless version in order
to understand how $V$ depends on the various length scales involved.
The form
\begin{equation}
    \tilde{V}(\tfrac{d}{y_0},\tfrac{d}{\xi},\alpha) =
    \left(\frac{d}{\sigma}\right)^2\frac{V}{B_0^2} = \int_0^{\infty}\tilde{S}\xi
    dq
    \label{Vtilde}
\end{equation}
renders the field variance dimensionless and makes it a function of
$d/y_0$, $d/\xi$ and $\alpha$. This is plotted versus $d/y_0$ in
figure \ref{Vtilde_fig} for three values of $d/\xi$ and two values
of $\alpha$. We see immediately that the dimensionless variance
$\tilde{V}$ is approximately constant with distance, corresponding
to a $1/d^4$ variation in $V$ that weakens when $d\lesssim y_0$.
This result is consistent with experimental observations that the
noise decreases with increasing distance from the wire
\cite{Jones04,Zimmermann2,Schumm05}. All the dependence on $\xi$ and
$\alpha$ is contained in $\tilde{V}$, which is largest when
$d\simeq\xi$ (solid lines). Changing $\alpha$ from $\alpha=1$ (light
curves) to $\alpha=1/4$ (heavy curves) makes essentially no
difference when $\xi/d$ is small (dashed lines), because $\tilde{P}$
is effectively a constant under these conditions, as discussed
above. By contrast, the same change of $\alpha$ reduces the noise
when $d\simeq\xi$ (solid line) and increases it when $d\ll\xi$
(dotted line). Another view of the same parameter space is given in
figure \ref{Vtildexi}, which plots the field noise $\tilde{V}$ at
height $d=y_0$ versus $d/\xi$, for three values of $\alpha$. This
shows more clearly the peaking of the noise near $\xi=d$, and the
insensitivity to the value of the Hurst exponent when $\xi$ is
small.

The best present methods for fabricating atom chip wires yield
correlation lengths of order 1--100~nm, while current experiments
operate at distances in the range 1--100~$\mu$m. This places
experiments firmly in the domain of small $\xi/d$, where the value
of the Hurst exponent does not influence the roughness of the
magnetic trap significantly and we can take $\tilde{P}\simeq2/\pi$.
In this regime we find that $\tilde{V}\propto\xi/d$. For any
particular value of $y_0$ we can integrate equation (\ref{Vtilde})
numerically to obtain the constant of proportionality. For example,
with $y_0=d$ we find
\begin{equation}
    \tilde{V}(\tfrac{d}{y_0}=1,\tfrac{d}{\xi}\gg1,\alpha) \simeq
    0.274\;\frac{\xi}{d}\,.
    \label{Vtildelimit}
\end{equation}
For narrower wires, i.e. for $d>y_0$, this constant changes very
little as we have already seen in figure \ref{Vtilde_fig}.

\section{Consequences for the design of the atom trap}

One of the primary motivations for atom chips is to achieve small
traps with high trapping frequencies. This requires a high field
gradient, which is approximately $\mu_0 I/(2\pi d^2)$ for narrow
wires (by which we mean $d\lesssim y_0$). The tightest confinement
is achieved by bringing the atoms close to narrow wires, even though
smaller wires carry less current~\cite{Schumm05}, but this also
increases the roughness of the magnetic field. In many experiments,
the field has a maximum permissible variance, let us call it
$V_{max}$. For example, a Bose--Einstein condensate will break into
separate clouds unless the noise in the trapping potential is
smaller than the chemical potential. In these cases, the limit on
trap roughness imposes a minimum distance from the surface and hence
a maximum achievable magnetic field gradient.

The heat generated by electrical resistance limits the current that
can be tolerated in a lithographically fabricated wire to
$I_{max}=\kappa y_0 \sqrt{x_0}$ \cite{Groth04}, where $x_0$ is the
thickness of the wire (figure \ref{wire}) and the constant $\kappa$
characterises the heat flow across the interface between the wire
and the substrate. When the maximum current is passed through a wire
having $y_0\simeq d$, equations (\ref{Vtilde}) and
(\ref{Vtildelimit}) give the variance of $B_z$ as
\begin{equation}
    V = 0.274 \left(\frac{\sigma^2\xi}{d^3}\right) \left(\frac{\mu_0\kappa}{2\pi}\right)^2
    x_0\,.
    \label{VwithImax}
\end{equation}
Upon setting this equal to $V_{max}$, we find that the distance of
closest approach and the maximum field gradient are
\begin{eqnarray}
    d_{min} & = & \left[\,0.274\;\sigma^2\xi\left(\frac{\mu_0\kappa}{2\pi
    }\right)^2 \frac{x_0}{V_{max}}\right]^{1/3}\,,\label{dmin}\\
    B^\prime_{max} & = & \frac{\mu_0 I_{max}}{2\pi d_{min}^{\,2}}
    =\left(\frac{\mu_0\kappa\sqrt{x_0} \,V_{max}}{0.274\times 2\pi
    \sigma^2\xi}\right)^{1/3}\,.\label{Bprimemax}\\\nonumber
\end{eqnarray}
Let us take $\sqrt{V_{max}}=1$~mG, for which the rms roughness in
the potential corresponds to a temperature of 67~nK. We take
$\kappa\simeq 3\times10^7$~A m$^{-3/2}$~ \cite{Groth04,Schumm05},
which is typical for gold wires on a $\mbox{Si/SiO}_2$ substrate. We
further assume the values $x_0\simeq 1~\mu$m, $\sigma\simeq3$~nm,
and $\xi\simeq 20$~nm, which are typical of our present wires, as
discussed in section \ref{description}. Then the minimum distance of
approach to the wire is $d_{min}\simeq 6~\mu$m, the current in the
wire is $I_{max}\simeq170$~mA, and the corresponding maximum field
gradient is $B^\prime_{max}\simeq11$~T\,cm$^{-1}$.

In the presence of a bias field $B_z$ along the $z$-direction
(figure \ref{wire}), the potential energy of the trapped atom near
its equilibrium position is
$\mu_z(B_z^2+(B^\prime_{max}\rho)^2)^{1/2}$, where $\rho$ is the
transverse ($x y$) displacement. The corresponding frequency for
transverse harmonic oscillations is
\begin{equation}
    f_{max} = \frac{1}{2\pi}B^\prime_{max}\sqrt{\frac{\mu_z }{m
    B_z}},\vspace{6pt}
\end{equation}
where $m$ is the mass of the atom. Taking a typical value of
$B_z\simeq0.5$~G, the maximum transverse frequency for $^{87}$Rb
atoms in the $F=2$, $m_F=+2$ ground state is $f_{max}\simeq190$~kHz.
This result indicates that atom chips can achieve very high trapping
frequencies, comparable with those already demonstrated in optical
lattices, while remaining adequately smooth. The correspondingly
small extension of the vibrational ground state wavepacket is only,
17~nm, making such traps very promising for studing the physics of
1-dimensional cold gases~\cite{Pricoupenko04}.

\section{Discussion }
We have examined how a magnetic trap formed by a current-carrying
wire is sensitive to the roughness on the edges of the wire. In
particular, we have extended previous discussions to consider the
case of a self-affine fractal roughness spectrum with a correlation
length $\xi$. This is of interest because the methods used to
fabricate wires on an atom chip generally produce such roughness.
Our analysis has shown how the spectrum of trap roughness involves
an interplay between the spectrum of the wire roughness and the
spectrum of the transfer function that converts deviations of the
current into fluctuations of the magnetic field. In most current
experiments, there is a clear hierarchy of length scales in which
$d\simeq y_0\gg\xi$. When there is also a maximum acceptable
roughness of the magnetic trap, this leads to equation (\ref{dmin})
for the minimum operating distance between the trap and the wire.
There is correspondingly a maximum achievable field gradient given
by equation (\ref{Bprimemax}). These results argue for minimising
the quantity $\sigma^2\xi$ because this determines the spectral
density of the edge roughness at low frequency, which is what
generates the noise in the magnetic trap. In these cases where
$d\simeq y_0\gg\xi$, the Hurst exponent is less important because it
only affects the spectrum at frequencies above $1/\xi$, which do not
contribute significantly to the trap roughness. Naturally, if there
is a second power-law regime at long wavelength, as in some
wires~\cite{Schumm05}, then the corresponding Hurst exponent
significantly affects the magnetic trap roughness at that length
scale.

There is considerable interest in trapping atoms much closer to the
surface. Atoms trapped at sub-micron distances would begin to probe
the details of the short-scale noise and in that case the value of
$\alpha$ would be significant. In the range of 0.1--1~$\mu$m one
could intentionally create surfaces with a variety of noise spectra
in order to propagate BEC through custom-made disorder potentials as
a study of quantum localisation phenomena, as proposed by
\cite{Wang04}. Experiments of this type have already been done using
random optical potentials~\cite{Lye05,Schulte05,Clement05}, but this
magnetic disorder offers a different noise spectrum with the
possibility of very short correlation length. The effective
amplitude of the noise can be conveniently controlled by appropriate
modulations of the currents that form the trap, as recently
demonstrated by \cite{Trebbia2007}. A natural lower limit to the
distance of closest approach is set by the electromagnetic
attraction of the atom towards the surface --- the Van der Waals
(Casimir-Polder) force --- which grows as $1/d^4$ ($1/d^5$) and
overwhelms the trapping force at a distance of order
100~nm~\cite{Hinds91,Hinds94}. This force itself is also of
fundamental interest and can be measured further away from the
surface by means of cold atoms~\cite{Hinds94,Obrecht07}. Even more
exotically, one can hope to measure the gravitational attraction at
short range, which might exhibit departures from the Newtonian law
as a result of extra dimensions~\cite{Dimopoulos03}. These
experiments also require careful control over the noise of the
surface. We note that if the distance to the surface becomes much
less than the width of the wire, the corrugation of the surface
\cite{Schumm05} and imperfections of the bulk \cite{Kruger2005} may
contribute significantly to the magnetic trap roughness.

In this study we have not considered the role of additional
technical noise at long correlation lengths, but such noise
certainly exists as a result of imperfections in the fabrication
process and can make a significant contribution to the roughness of
the magnetic trap at long wavelength. For example, a supposedly
straight line may be bent by optical aberrations during the
lithographic patterning. A bend of only $100~\mu$rad amounts to a
deviation of only 10~nm over a length of $100~\mu$m. Nevertheless,
in a bias field of $B_0=10$~G this would generate an appreciable
unwanted $B_z$ of 1~mG. There can also be a periodic wobble of
straight lines due to the imperfection of mechanical translation
stages. These kinds of defects are very hard to measure by standard
microscopy because they involve small transverse displacements over
length scales larger than the normal field of view in an SEM or AFM
microscope. In fact, this noise is best measured by the cold atoms
themselves through its effect on their density distribution in the
trap. Ref.~\cite{Kruger2005} has already noted that on certain
length scales, this method may provide a uniquely sensitive way to
probe the magnetic field near surfaces.

\begin{acknowledgments}
This work was supported by the European Commission through the Atom
Chips, Conquest and SCALA networks and by the UK through EPSRC and
Royal Society funding. We acknowledge valuable discussions with S.
Eriksson.
\end{acknowledgments}


\begin{thebibliography}{99}
\bibitem{Hinds2001}
E. A. Hinds, C. J. Vale, and  M. G. Boshier, Phys. Rev. Lett.
\textbf{86} 1462 (2001).

\bibitem{Fortagh98}
J. Fort\'agh, A. Grossmann and C. Zimmermann, Phys. Rev. Lett.
\textbf{81} 5310 (1998).

\bibitem{Denschlag99}
J. Denschlag, D. Cassettari  and J. Schmiedmayer, Phys. Rev. Lett.
\textbf{82} 2014 (1999).

\bibitem{Key2000}
M. Key, I. G. Hughes, W. Rooijakkers, B. E. Sauer, E. A. Hinds, D.
J. Richardson and P. G. Kazansky, Phys. Rev. Lett. \textbf{84} 1371
(2000).

\bibitem{Fortagh07}
J. Fort\'agh and C. Zimmermann, Rev. Mod. Phys. \textbf{79} 235
(2007).

\bibitem{Eriksson05}
S. Eriksson, M. Trupke, H. F. Powell, D. Sahagun, C. D. J. Sinclair,
E. A. Curtis, B. E. Sauer, E. A. Hinds, Z. Moktadir, C. O. Gollasch,
M. Kraft, Eur. Phys. J. D \textbf{35} 135 (2005).

\bibitem{Gollasch05}
C. O. Gollasch, Z. Moktadir, M. Kraft, M. Trupke, S. Eriksson and E.
A. Hinds, J. Micro. Mech. and Micro. Eng. \textbf{15} S39 (2005).

\bibitem{Calarco04}
C. Henkel, J. Schmiedmayer, and C. Westbrook, Euro. Phys. J. D
\textbf{35} 1 (2006) and following articles.

\bibitem{Jones03}
M. P. A. Jones, C. J. Vale, D. Sahagun, B. V. Hall and E. A. Hinds,
Phys. Rev. Lett. \textbf{91} 080401 (2003).

\bibitem{Zimmermann1}
J. Fort\'agh, H. Ott, S. Kraft, A. G\"{u}nther and C. Zimmermann,
Phys. Rev. A \textbf{66} 041604 (2002).

\bibitem{Jones04}
M. P. A. Jones, C. J. Vale, D. Sahagun, B. V. Hall, C. C. Eberlein,
B. E. Sauer, K. Furusawa, D. Richardson and E. A. Hinds , J. Phys. B
\textbf{37} L15 (2004).

\bibitem{Zimmermann2}
S. Kraft, A. G\"{u}nther, H. Ott, D. Wharam, C. Zimmermann and J.
Fort\'agh, J. Phys. B \textbf{35} L469 (2002).

\bibitem{Schumm05}
T. Schumm, J. Esteve, C. Figl, J.-B. Trebbia, C. Aussibal, H.
Nguyen, D. Mailly, I. Bouchoule, C.I. Westbrook and A. Aspect, Eur.
Phys. J. D \textbf{32} 171 (2005).

\bibitem{Wang04}
D. W. Wang, M. D. Lukin and E. Demler, Phys. Rev. Lett. \textbf{92}
076802 (2004).

\bibitem{Kraft04}
E. Koukharenko, Z. Moktadir, M. Kraft, M. E. Abdelsalam, D. M.
Bagnall, C. Vale, M. P. A. Jones and E. A. Hinds, Sensors and
Actuators A \textbf{115} 600 (2004).

\bibitem{Lewis06}
G. Lewis, Z. Moktadir, C. O. Gollasch, Kraft, M. Trupke, S.
Erikisson and E. A. Hinds, Proceedings of 16$^{th}$ MicroMechanics
Europe Workshop (2006).

\bibitem{Meakin}
P. Meakin, {\it Fractals, scaling and growth far from equilibrium}
(Cambridge University Press, 1998).

\bibitem{Barabasi}
A. L. Barabasi and H. E. Stanly, {\it Fractal concept in surface
growth} (Cambridge University Press, 1995).

\bibitem{Constantoudis03}
V. Constantoudis, G. P. Patsis, A. Tserepi and E. Gogolides, J. Vac.
Sci. Tech. B\textbf{21} 1019 (2003).

\bibitem{Palasantzas93}
G. Palasantzas, Phys. Rev. B \textbf{48} 14472 (1993).

\bibitem{note1}
When $\alpha>1$, the surface is said to be super rough. Films grown
by molecular beam epitaxy, where surface diffusion is the dominant
process, are one example.

\bibitem{Constantoudis04}
V. Constantoudis, G. P. Patsis, L. H. A. Leunissen and E. Gogolides,
J. Vac. Sci. Tech. B \textbf{22} 1974 (2004).

\bibitem{Groth04}
S. Groth, P. Kr\"{u}ger, S. Wildermuth, R. Folman, T. Fernholz, J.
Schmiedmayer, D. Mahalu and I. Bar-Joseph, Appl. Phys. Lett.
\textbf{85} 2980 (2004).

\bibitem{Pricoupenko04}
{\it Quantum gases in low dimensions}, edited by L. Pricoupenko, H.
Perrin, and M. Olshanii, special issue of J. Phys. IV 116 1 (2004).

\bibitem{Lye05}
J. E. Lye, L. Fallani, M. Modugno, D. S. Wiersma, C. Fort, and M.
Inguscio, Phys. Rev. Lett. \textbf{95} 070401 (2005).

\bibitem{Schulte05}
T. Schulte, S. Drenkelforth, J. Kruse, W. Ertmer, J. Arlt, K. Sacha,
J. Zakrzewski, and M. Lewenstein, Phys. Rev. Lett. \textbf{95}
170411 (2005).

\bibitem{Clement05}
D. Cl\'{e}ment, A. F. Var\'{o}n, M. Hugbart, J. A. Retter, P.
Bouyer, L. Sanchez-Palencia, D. M. Gangardt, G.V. Shlyapnikov, and
A. Aspect, Phys. Rev. Lett. \textbf{95} 170409 (2005).

\bibitem{Trebbia2007}
J.-B. Trebbia, C.L. Garrido Alzar, R. Cornelussen, C.I. Westbrook
and I. Bouchoule, arXiv:quant-ph/0701207 (2006).

\bibitem{Hinds91}
E. A. Hinds and V. Sandoghdar, Phys. Rev. A \textbf{43} 398 (1991).

\bibitem{Hinds94}
E. A. Hinds, Adv. At. Mol. Opt. Phys., Supplement 2, edited by Paul
R. Berman (Academic Press, Inc., 1994), p.~1.

\bibitem{Obrecht07}
J.M. Obrecht, R.J. Wild, M. Antezza, L. P. Pitaevskii, S. Stringari,
and E. A. Cornell, Phys. Rev. Lett. \textbf{99} 063201 (2007).

\bibitem{Dimopoulos03}
S. Dimopoulos and A. A. Geraci, Phys. Rev. D \textbf{68} 124021
(2003).

\bibitem{Kruger2005}
P. Kr\"{u}ger, S. Wildermuth, S. Hofferberth, L. M. Andersson, S.
Groth, I. Bar-Joseph and J. Schmiedmayer, J. of Phys.: Conf. Series
\textbf{19} 56 (2005).
\end{thebibliography}
\end{document}